# Single-Pilot-Tone-Enabled Low-Complexity Full-Band Clock Recovery for Coherent DSCM Systems

**Wei Wang,[1,2] Dongdong Zou,[1] Mengqi Wang,[1] Jiawen Yao,[1] and Fan Li[2,*]**
[1] *School of Electronic and Information Engineering, Soochow University, Suzhou 215006, China*
[2] *School of Electronics and Information Technology, State Key Laboratory of Optoelectronic Materials and Technologies, Sun Yat-Sen University, Guangzhou 510275, China*
**lifan39@mail.sysu.edu.cn*

**Abstract:** This paper proposes a low-complexity single-pilot-tone-enabled full-band clock recovery scheme for coherent DSCM systems. Pilot-tone-based timing error estimation enables clock recovery before subcarrier demultiplexing, eliminating redundant per-subcarrier CR loops and providing a compact Rx-DSP architecture.

## 1. Introduction

The rapid growth of AI-driven data traffic is fueling the demand for higher transmission capacities in optical interconnects. To meet these demands, coherent technology is also evolving towards higher baud rates and improved spectral efficiency. In this case, the single-carrier signal with increased baud rate and modulation order becomes increasingly susceptible to system impairments, posing a significant challenge to improving the transmission capacity [1]. Coherent digital subcarrier multiplexing (DSCM) technique has emerged as a promising candidate by partitioning a high baud-rate signal into multiple low baud-rate subcarriers, offering advantages in chromatic dispersion compensation, tolerance to fiber nonlinearity, and flexible subcarrier allocation [2].

Despite these advantages, the implementation complexity of receiver-side digital signal processing (Rx-DSP) remains a critical concern. In contrast to the DSP chain of conventional single-carrier systems, each subcarrier should be operated independently after subcarrier demultiplexing, including several typical DSP modules such as clock recovery (CR), equalization, and carrier phase recovery (CPR). Such a repetitive processing framework increases hardware resource consumption, particularly as the number of subcarriers grows [3]. Therefore, performing DSP operations jointly across multiple subcarriers could significantly reduce redundant processing and alleviate receiver-side DSP complexity [4]. In our previous work, we successfully demonstrated polarization demultiplexing of the entire DSCM signal prior to subcarrier demultiplexing using a single frequency-domain pilot tone, thereby substantially alleviating the complexity of the equalizer required for each subcarrier [5]. Moreover, clock recovery is also a crucial DSP function for reconstructing the received signal at the optimal sampling instants, while the commonly used timing error detectors (TEDs), such as the Godard phase detector (GPD), Gardner algorithm, and absolute phase detector (ABSPD), rely on the intrinsic characteristics of the signal for timing error estimation [6-8]. Consequently, reliable timing error estimation directly from the composite DSCM signal remains challenging, necessitating an independent CR loop for each subcarrier. This leads to considerable hardware redundancy and increased implementation complexity.

To address this limitation, a low-complexity full-band clock recovery scheme for coherent DSCM systems using a single frequency-domain cosine pilot tone is proposed in this paper. By deriving timing error information from the pilot tone rather than the data-bearing subcarriers, the proposed scheme performs clock recovery on the composite DSCM signal prior to subcarrier demultiplexing. It replaces multiple independent CR loops with a shared full-band operation, thereby reducing redundant hardware resources and improving the compactness of Rx-DSP. Furthermore, the pilot tone can be reused for frequency offset estimation (FOE) and compensation, avoiding additional pilot overhead. Experimental results based on a 35Gbaud 4-subcarrier 16QAM (4SC-16QAM) coherent DSCM system demonstrate that the proposed scheme achieves comparable BER performance to conventional independent CR loop approaches, while enabling a more streamlined Rx-DSP architecture.

## 2. Principle

Figures 1(a) and 1(b) show the conventional and proposed Rx-DSP architectures of the coherent DSCM system. In contrast to the conventional one, the proposed scheme derives a common timing error from a single pilot tone and corrects the sampling instant of the composite DSCM signal prior to subcarrier demultiplexing. Consequently, the separate DSP chain for each subcarrier can be pruned. In the proposed scheme, the cosine pilot tone is inserted at X-polarization between two subcarriers as illustrated in Fig. 1(c). The pilot is also shared with the FOE module. When the sampling rate is four times the pilot-tone frequency, the sampled pilot tone exhibits a twofold oversampling characteristic, as shown by the red dot in Figs. 1(d) and 1(e). In this case, the Gardner algorithm can be implemented to obtain the timing error as:

$$\varepsilon = s_{2n} \cdot \left(s_{2n-1} - s_{2n+1}\right) \tag{1}$$

where $s_n$ is the sampled pilot tone. According to Figs. 1(d) and 1(e), if sampling occurs earlier than the optimal sampling instant (green dot), the resulting timing error is positive; conversely, late sampling (blue dot) produces a negative timing error. For a hardware-efficient implementation, equation (1) can be reformulated using sign operations as:

$$\varepsilon = \mathrm{sgn}\left(s_{2n}\right) \cdot \mathrm{sgn}\left(s_{2n-1} - s_{2n+1}\right) \tag{2}$$

where $\mathrm{sgn}(\cdot)$ refers the sign operation. This modification preserves the polarity of the timing error while effectively eliminating multiplication operations.

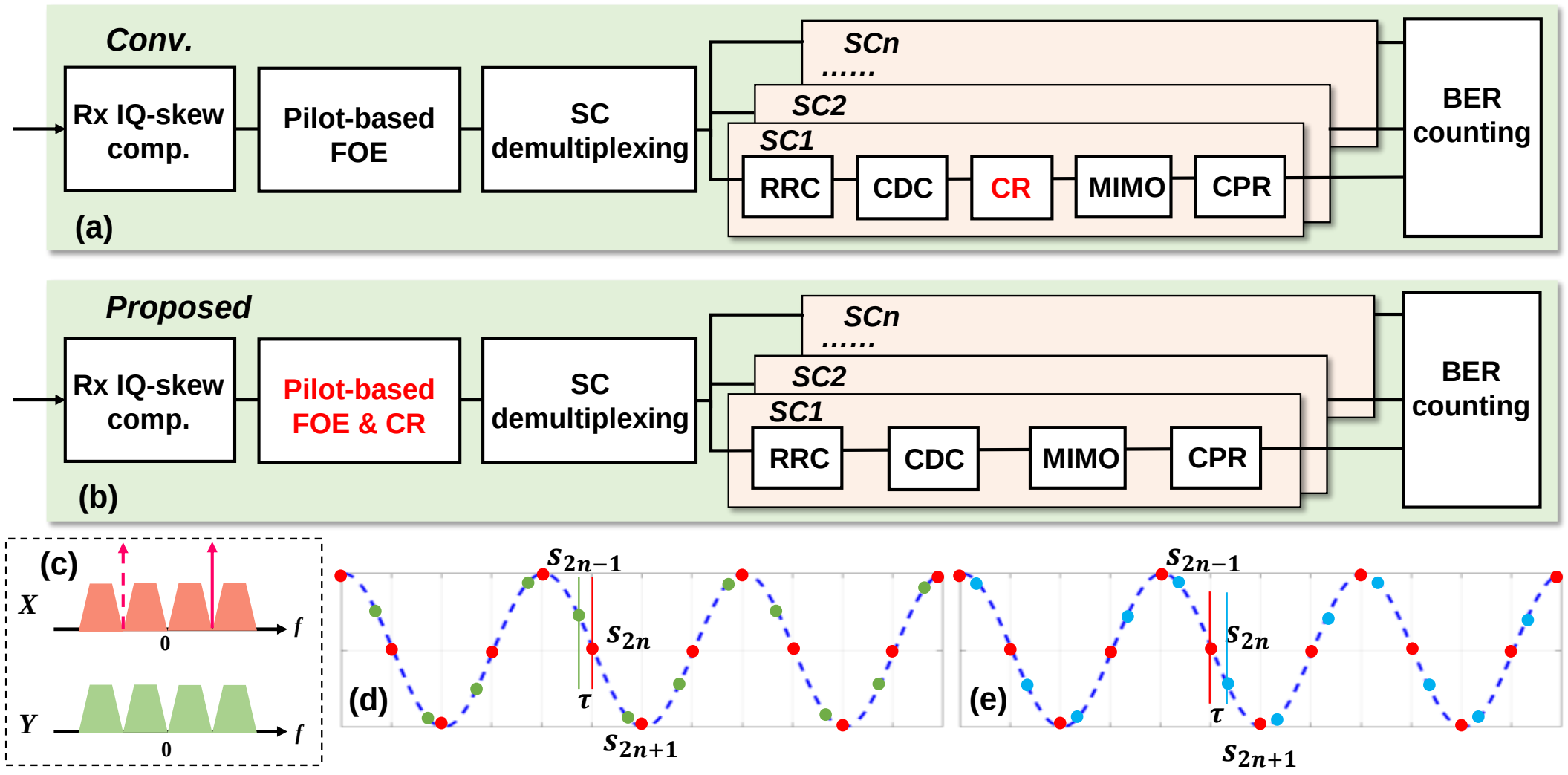

Fig. 1. (a) Conventional Rx-DSP flow. (b) Proposed Rx-DSP flow. (c) Schematic spectrum of the DSCM signal with an inserted pilot tone. The sampled pilot tone signal in the time domain: (d) early sampling, (e) late sampling.

## 3. Experimental Setup and Results

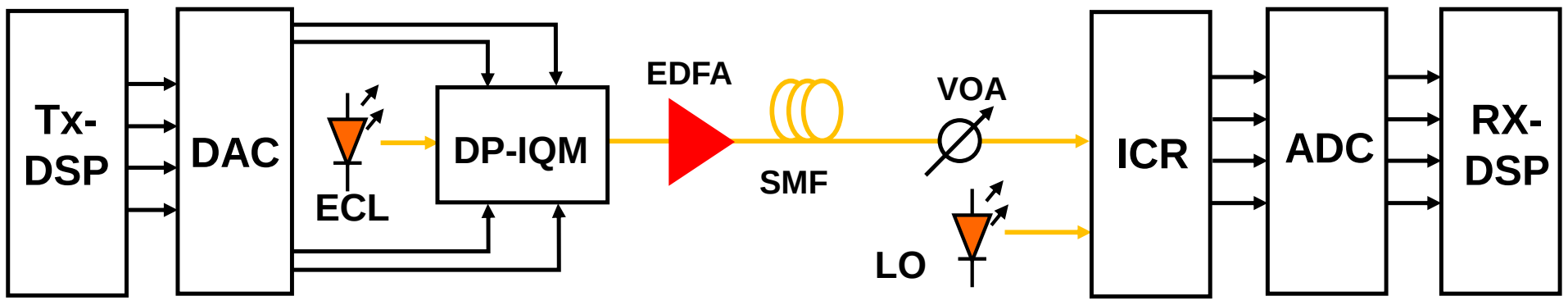

Fig. 2. Experimental setup.

To evaluate the effectiveness of the proposed single-pilot-tone-enabled full-band clock recovery scheme, a 35 Gbaud 4SC-16QAM coherent DSCM system is established as shown in Fig. 2. At the transmitter side, the DSCM signal with an inserted pilot tone is generated offline and uploaded into the Fujitsu digital-to-analog converter (DAC) with a sampling rate of 80 GSa/s. After digital-to-analog conversion, the electrical signals are loaded into an integrated dual-polarization IQ modulator (DP-IQM). The optical carrier is provided by an external cavity laser (ECL) with a linewidth less than 100 kHz and output power of 13 dBm. Then, the modulated signal is amplified by an Erbium-doped fiber amplifier (EDFA) before being injected into the fiber link to adjust the launch power. At the receiver side, a variable optical attenuator (VOA) is implemented to adjust the received optical power (ROP). Then, the signal is mixed with a 13 dBm local oscillator (LO) in an integrated coherent receiver (ICR), and the detected signals are sampled by a Fujitsu analog-to-digital converter (ADC) operating at 80 GSa/s. Finally, the captured signals are processed in offline DSP. The two architectures of Rx-DSP are given in Figs. 1(a) and 1(b). As for the Tx-DSP, the pseudo-random binary sequence (PRBS) is mapped into 16QAM symbols for each subcarrier. After being shaped by a root-raised cosine (RRC) filter with a roll-off factor of 0.1, four subcarriers are multiplexed to form a 4SC-16QAM DSCM signal for each polarization. Then, the whole signal is resampled to match the sampling rate of the DAC. After that, a cosine pilot tone is inserted into the DSCM signal in the X-polarization. Finally, the transmitter-side IQ skew (Tx-skew) is calibrated before digital-to-analog conversion.

Initially, the pilot-to-signal power ratio (PSPR) is optimized under different sampling frequency offset (SFO) conditions in the optical back-to-back (OBTB) configuration. The experimental results are shown in Fig. 3(a), with the PSPR changing from -30.7 dB to -1.2 dB. The bit error rate (BER) performance under different SFO conditions follows a similar trend, and the optimal PSPR is about -16.7 dB. Therefore, the PSPR is fixed at -16.7 dB in subsequent measurements. Fig. 3(b) shows the BER curves of the proposed full-band clock recovery scheme versus optical signal-to-noise ratio (OSNR). The experimental results show that nearly identical BER performance is achieved across different SFO conditions, and it also matches the case without SFO, indicating the robustness of the proposed scheme against SFO. Furthermore, the BER performance of the proposed scheme is also compared to the independent clock recovery schemes based on several commonly used TED schemes, including ABSPD, GPD, and the Gardner algorithm. Fig. 3(c) compares the BER performance of these four different clock recovery schemes at the OSNR of 29dB. It can be observed that, with the increase of the SFO, the BER of the signal with an independent ABSPD-based CR loop for each subcarrier will be slightly worse, while the BER curve of the proposed scheme shows no noticeable performance degradation. Additionally, the performance over fiber transmission is also investigated. Fig. 3(d) gives the BER curves versus ROP of different clock recovery schemes over 80 km single-mode fiber (SMF) transmission. The BER curves of the signal using the proposed full-band clock recovery scheme and independent CR loop based on GPD and Gardner algorithm nearly overlap, while the performance of the independent ABSPD-based CR loop will be slightly inferior, particularly at an SFO of 90 ppm.

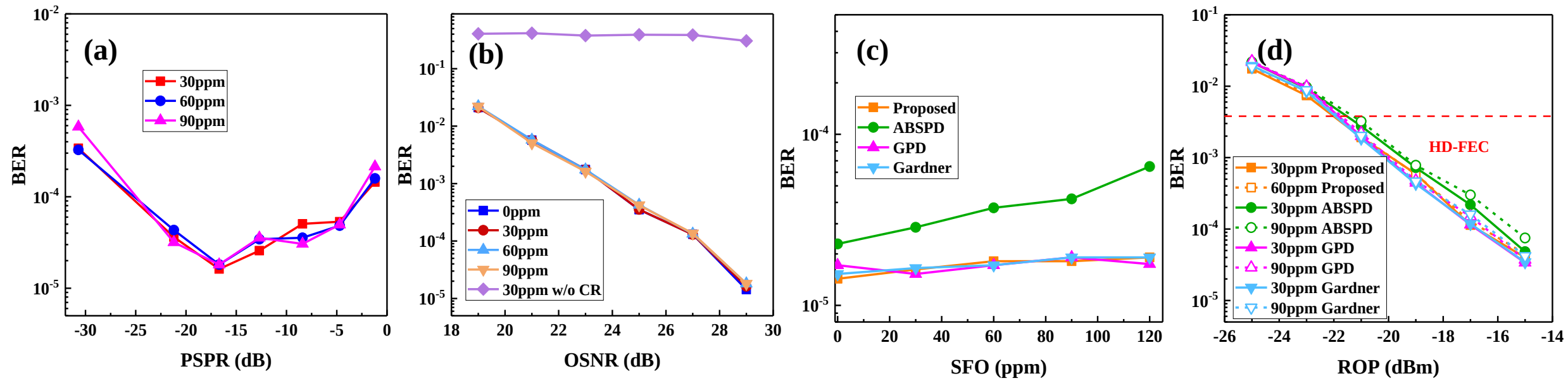

Fig. 3. (a) BER versus PSPR of the proposed scheme in OBTB. (b) BER versus OSNR of the proposed scheme in OBTB. (c) Performance comparison of several clock recovery schemes under different SFO conditions in OBTB. (d) BER versus ROP of different clock recovery schemes in 80 km SMF transmission.

## 4. Conclusion

This work proposes a novel clock recovery structure for coherent DSCM systems. By extracting timing error from the pilot tone, the proposed scheme achieves full-band clock recovery before subcarrier demultiplexing, eliminating redundant per-subcarrier CR loops. Experimental results demonstrate comparable BER performance to conventional schemes, while enabling a more compact and hardware-efficient Rx-DSP architecture.

## 5. Acknowledge

This work is partly supported by the National Key R&D Program of China (2023YFB2906000); National Natural Science Foundation of China (62271517, U2001601, 62035018); Guangdong Basic and Applied Basic Research Foundation (2023B1515020003), State Key Laboratory of Advanced Optical Communication Systems and Networks of China (2024GZKF19); Postdoctoral Fellowship Program of China Postdoctoral Science Foundation (GZB20250813); Jiangsu Founding Program for Excellent Postdoctoral Talent (2025ZB335).

## 6. References

[1] Y. Jin, et al., JLT, 37(14), 769-772, 2025.
[2] Y. Pan, et al., JLT, 44(20), 8952-8965, 2026.
[3] L. Fan, et al., JLT, 41(5), 1454-1463, 2023.
[4] L. Fan, et al., JLT, 41(19), 6187-6198, 2023.
[5] W. Wang, et al., JLT, 43(3), 1150-1161, 2025.
[6] C. Li, et al., OE, 31(11), 18599-18612, 2023.
[7] N. Stojanovic, et al., OFC 2020, Paper M4J.5.
[8] W. Wang, et al., OECC 2025, Paper WB3-4.